# A fast, sensitive, room-temperature graphene nanomechanical bolometer

Andrew Blaikie, David Miller, Benjamín J. Alemán

Department of Physics, University of Oregon, Eugene, Oregon 97403
Material Science Institute, University of Oregon, Eugene, Oregon 97403
Center for Optical, Molecular, and Quantum Science, University of Oregon, Eugene, Oregon 97403

**Abstract**

**Bolometers are a powerful and vital means of detecting light in the IR to THz frequencies, and they have been adopted for a range of uses from astronomical observation to thermal imaging. As uses diversify, there is an increasing demand for faster, more sensitive room-temperature bolometers. To this end, graphene has generated interest because of its miniscule heat capacity and its intrinsic ultra-broadband absorption, properties that would allow it to quickly detect low levels of light of nearly any wavelength. Yet, graphene has disappointed its expectations in traditional electrical bolometry at room temperature, because of its weakly temperature-dependent resistivity and exceptionally high thermal conductivity. Here, we overcome these challenges with a new approach that detects light by tracking the resonance frequency of a graphene nanomechanical resonator. The absorbed light heats up and thermally tensions the resonator, thereby changing its frequency. Using this approach, we achieve a room-temperature noise-equivalent power of 7 pW/Hz$^{1/2}$, a value 100 times more sensitive than electrical graphene bolometers, and speeds (1.3 MHz) that greatly surpass state-of-the-art microbolometers.**

Bolometers detect electromagnetic radiation and energetic particles by measuring an increase in temperature due to absorption of the radiation or the particle. In terms of radiation detection, the benefit of the bolometer is its ability to detect light deep into the infrared[1]. This ability has led to many applications and scientific discoveries[2], including thermal imaging and night vision, IR spectroscopy, and the mapping of the cosmic microwave background. In addition to the ability to absorb a broad spectrum of electromagnetic radiation, the ideal bolometric material must possess a large thermal responsivity, traditionally achieved via temperature-dependent electrical resistance. Also, this material would possess a low thermal conductivity and low thermal mass, which increase the temperature change for a given amount of absorbed energy and the response speed. Furthermore, emerging applications like remote and portable medical imaging[3] and environmental monitoring, security, communication, and earth and solar science[4] demand that the bolometer material be useful at and above room temperature. Graphene has attracted recent interest for next-generation bolometry[5,6] because of its ultra-low mass and its ability to efficiently absorb radiation from the ultraviolet to terahertz and radio frequencies[7,8]. In principle, graphene could serve as the basis for fast, sensitive, ultra-broadband bolometry. However, graphene's electrical resistivity has an weak temperature dependence[9], varying less than 10 % across a range of 1.6-300 K, and graphene's thermal conductivity, instead of being low as desired, is among the highest of any material[10]. Together, these properties yield a poor responsivity[11] which has limited the performance of graphene in traditional resistive bolometry at room temperature[6]. Departures from the traditional approach have been promising. For example, several implementations of graphene bolometry have been achieved by measuring the electron temperature rather than the lattice temperature[12–15]. However, these approaches rely on weak electron-phonon interactions and a low electronic thermal conductivity that occur at cryogenic temperatures[16]. Therefore, at room temperature, these approaches become markedly compromised or do not work.

Here, we report a new type of graphene bolometer, one that abandons the electronic resistive mechanism altogether, and instead employs nanoscale thermomechanics. We detect absorbed light by tracking the resonance frequency of a graphene nanomechanical resonator. The absorbed light heats up and tensions the resonator by thermal contraction, thereby changing its mechanical resonance frequency (Figure 1a). Similar nanomechanical-based resonant sensing has been demonstrated to detect several quantities in the single-quanta limit, including the proton mass[17] and the charge and spin of the electron[18,19]. Here, we extend the success of nanomechanical sensing to the detection of light.

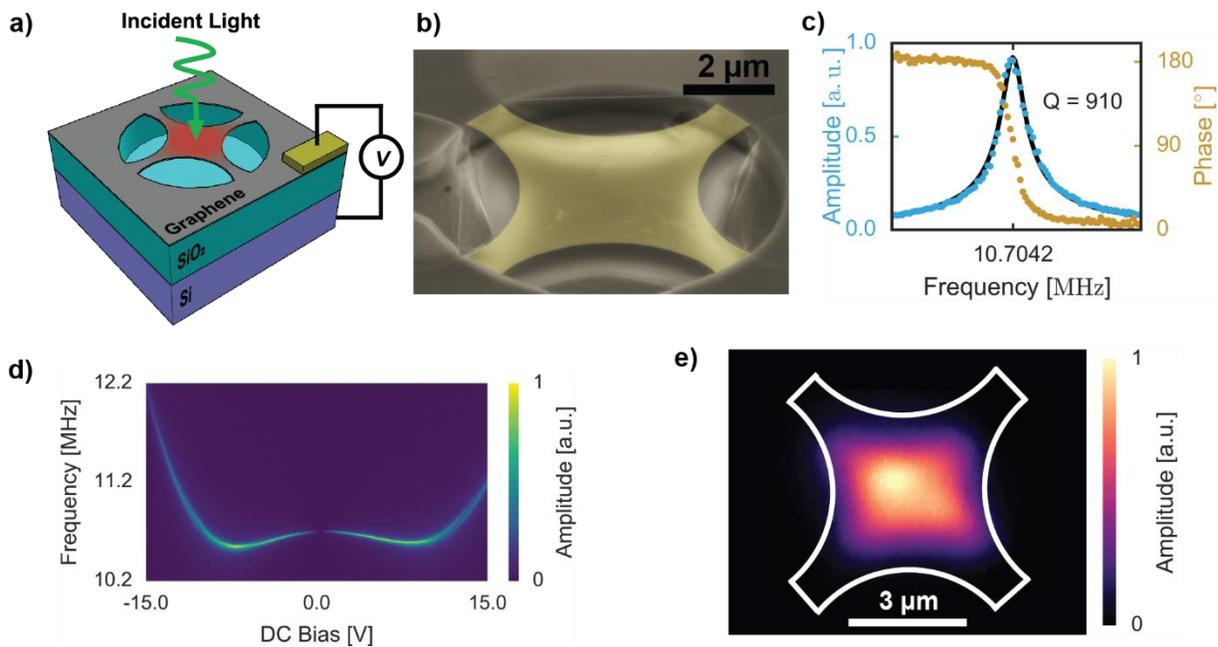

Figure 1: **a)** Illustration of the bolometric detection scheme. A driving voltage, $V_{AC}$, is used to actuate motion and a bias voltage, $V_{DC}$, is used to apply additional tension. The total voltage across the device is $V = V_{DC} + V_{AC}$. Absorbed light tightens the graphene, shifting the mechanical resonance. **b)** False-color scanning-electron-microscope image of a suspended graphene trampoline. Regions of collapsed graphene from the focused ion beam cutting process can be seen around the edges of the cavity. **c)** Amplitude-frequency response curve at $V_{DC} = 0.25$ V. The frequency of $V_{AC}$ swept as the mechanical amplitude response is measured. A best fit for a damped driven oscillator is used to calculate the resonance frequency and quality factor. **d)** Amplitude response vs. applied bias and drive-frequency spectrogram. The linewidth is observed to broaden with increasing $V_{DC}$. **e)** Measured mechanical mode shape of a graphene trampoline. Fast steering mirrors were used to scan the probe laser across the device with diffraction limited resolution. The white lines are outline the physical device shape as calibrated from a scanning-electron-microscope image.

The key components of our graphene nanomechanical bolometer (GNB) are a low heat capacity, a large thermal resistance, and a mechanical frequency that, in contrast to the electrical resistance of graphene, possesses a strong temperature dependence[20–22]. To illustrate this dependence, we calculate the temperature change required to shift the frequency of our graphene resonator by an amount $\Delta f_0$,

$$\Delta T \approx \left(\frac{4\pi\sigma_0(1-\nu)}{\alpha Y}\right)\frac{\Delta f_0}{f_0} \quad (1)$$

where $f_0$ is the initial frequency (also, $\sigma_0$ is the initial in plane stress of the graphene sheet, $\alpha$ is the thermal expansion coefficient, $\nu$ is the Poisson ratio, and $Y$ is the elastic modulus.) Using typical values for the material properties of graphene resonators[23–25] (see SI for details), we obtain $\Delta T \sim 300$ mK for $\Delta f_0$ equal to a full line-width, or, using expected values for the fractional frequency noise resolution, we obtain a minimal detectable temperature of order ~ 1-10 mK. In our system, the $\Delta T$ for a given absorbed energy is amplified by the ultra-low heat capacity and abnormally large thermal resistance[26] of suspended graphene, and we increase $\Delta T$ further by shaping the suspended graphene into a trampoline geometry (see Fig. 1b). The narrow, tapered tethers of the trampoline increase the thermal resistance between the center of the trampoline and the surrounding substrate support, which also acts as a thermal sink. Also, the miniscule heat capacity makes our system reach steady-state heat flow very quickly, resulting in a fast response bandwidth. Altogether, our GNB achieve a record room-temperature sensitivity (~ 7 pW/Hz$^{1/2}$) and a response bandwidth (~ 1 MHz) nearly a million-times faster than modern commercial bolometers. The physical size and scalable fabrication of our graphene detector could enable high-density bolometer arrays.

**The graphene nanomechanical bolometer and mechanical response**

The basic light-sensing component of our GNB is a graphene drumhead, which we fabricate by suspending graphene over circular holes in silicon oxide. The structures are made using standard semiconductor processing techniques and a dry polymer-supported graphene transfer technique[27] (see Methods). We shape drumheads into trampoline geometries using focused ion beam milling (FIB)[28]. Figure 1b shows a scanning-electron-microscope image of a graphene trampoline (shaded in yellow) with an 8 µm diameter and 500 nm wide tethers. We actuate mechanical motion of the graphene by applying an AC ($V_{AC}$) and DC ($V_{DC}$) bias voltage between a silicon backgate and the graphene, and we detect its motion with a scanning laser interferometer (633 nm, < 1 µW) by employing the graphene and silicon as a Fabry-Pérot cavity. For bolometric photodetection measurements, we use an additional laser (532 nm) as a heating source, which we modulate with an acousto-optic modulator (AOM). The resulting heat-induced shifts to the mechanical resonance frequency are monitored with a phase-locked loop (PLL).

To detect light, our nanomechanical bolometer relies on the measurement of the mechanical resonance frequency. By sweeping the frequency of $V_{AC}$, we obtain amplitude and phase spectra as seen in Figure 1c for the first fundamental mode. The frequency of this resonator is $f_0 = 10.7$ MHz with a $Q = 910$, or linewidth of 11.8 kHz. These spectra and values agree with the response of a graphene mechanical resonator of these dimensions. As further confirmation of mechanical resonance[29], we use the scanning interferometer to obtain a two-dimensional spatial map of the vibrational amplitude of a trampoline driven on resonance (Figure 1e). The map has four lobes, corresponding to each of the four tethers, and constant phase, consistent with an out-of-plane fundamental vibrational mode. Due to capacitive tensioning, the resonance frequency is sensitive to $V_{DC}$, as shown in the spectrogram (Figure 1d), which also shows no detectable amplitude at $V_{DC} = 0$ V, as expected for our actuation scheme[30]. Because of resistive dissipation[31], the quality factor ($Q$) decreases with $V_{DC}$, as can be seen from the linewidth broadening in the spectrogram. To maximize the signal-to-noise ratio and to optimize the bolometric responsivity, we set $V_{DC} \leq$ 1 V and $V_{AC}$ just below the onset of Duffing nonlinearity, typically less than 100 mV.

**Thermomechanical photoresponse**

The power sensitivity of our GNB is governed in part by the frequency-shift responsivity (*i.e.* the fractional change in resonance frequency per unit of absorbed power), given by

$$R_f \equiv \frac{\Delta f_0}{f_0 I} \tag{2}$$

where $\Delta f_0$ is the change in resonance frequency due to absorption of radiation at power $I$, and $f_0$ is the initial resonance frequency. To determine $R_f$, we track $\Delta f_0$ with a PLL while we sinusoidally modulate the power of the heating laser. A temporal measurement of $\Delta f_0$ with incident power $I_p = 190$ nW modulated at 40 Hz is shown in Figure 2a. We see $\Delta f_0$ matches the sinusoidal heating and has a peak-to-peak amplitude $\Delta f_0 = 8.5$ kHz as determined through fitting to a sine wave (black trace); this shift is ~ 72% of the resonator linewidth and is noticeably far above the background noise. To ascertain the power dependence, we measure $\Delta f_0$ for varying laser powers up to 5 µW, as shown in Figure 2b, and find a uniform linear resonance shift per unit incident power of $\Delta f_0/I_p = 52.5$ kHz/µW. Ultimately, we varied the range of incident power from about 50 nW to over 500 µW with no observable degradation in the response, thus demonstrating a dynamic range spanning four orders of magnitude and a shift in $f_0$ of over 500%. Using absorbed power, here taken to be at 2.3%[7,8] of incident, we measured $R_f$ for 9 different trampolines and 3 different drumheads. We plot $R_f$ against tether width ($w$) in Figure 2c. The trampoline width ($w$) and diameter ($d$) are indicated in Figure 2e. In general, the drumheads, which lack the added resistance of tethers, had $R_f$ values about 1% that of trampolines and dropped as low as 2,600 W$^{-1}$. For trampolines, $R_f$ increases for larger diameter and for narrower tether width. Our most sensitive device, a 6 µm diameter trampoline with 200 nm wide tethers, had $R_f \sim 300{,}000$ W$^{-1}$, a figure that is a factor 100 greater than state-of-the-art resonant microbolometers[32–34].

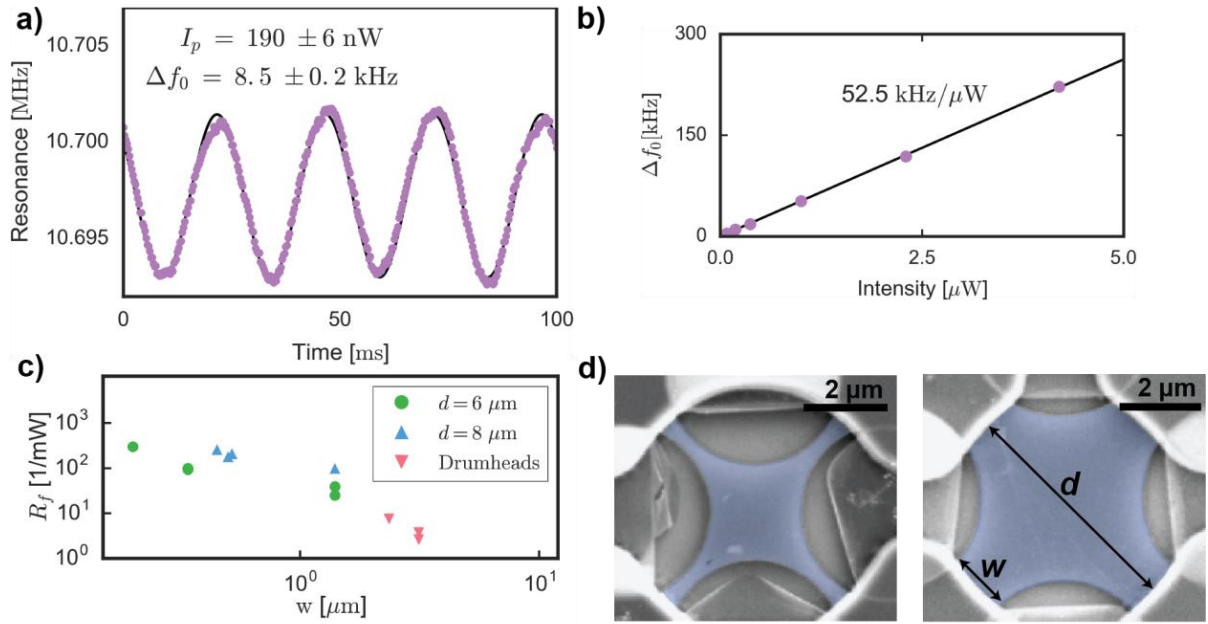

Figure 2: **a)** Mechanical resonance frequency vs. time as 190 nW incident radiation is modulated at 40 Hz. A phase-locked loop is used to temporally track changes in resonance frequency. **b)** Measured resonance shift as the incident heating laser intensity is increased. A best-fit line to this data yields a 52.5 kHz/µW resonance shift per incident power. **c)** Frequency responsivity vs. tether width, *w*, for 9 different trampolines and 3 different drumheads. For the drumheads, the tether width is taken to be 1/4 of the drumhead diameter. **d)** False colored scanning electron microscope images of two different trampolines with different tether widths, *w*, with diameter, $d = 6$ µm.

**Noise-equivalent power and response speed**

Together with the frequency shift responsivity, the fluctuations of the resonance frequency ($\sigma_f$) determine the noise-equivalent power per Hz$^{1/2}$ ($\eta$) of our GNB through the expression

$$\eta = \frac{\sigma_f \sqrt{t}}{f_0 \, R_f} \tag{3}$$

where $t$ is the measurement time. A routine measure of the fractional noise, $\sigma_f/f_0$, is the Allan deviation[35], defined as

$$\sigma_A^2 = \frac{1}{2(N-1)f_0^2} \sum_{m=2}^{N} (f_m - f_{m-1})^2, \tag{4}$$

where $f_m$ is the average frequency measured over the $m$th time interval of length $t$. We measure the frequency time traces with a PLL while the heating laser is turned off, and then we calculate $\sigma_A$, as shown in Figure 3a, for sampling intervals ranging from 10 ms to 1 s. Across this sampling range, the Allan deviation for all devices, including both trampolines and drumheads, remains nearly constant around $\sigma_A \sim 10^{-5}$. This flat response indicates that the mechanical frequency noise is dominated by flicker noise ($1/f$), which is consistent with other resonators of similar mass[36].

The most clear geometric dependence of $\eta$ is with tether width—wider tethers lead to poorer sensitivity (Figure 3b). A trampoline with 200-nm-wide tethers, the narrowest tether width tested, possessed the lowest noise-equivalent power, $\eta = 7$ pW/Hz$^{1/2}$. For drumheads, which lack tethers, $\eta$ is over 200 times less sensitive than trampolines, with $\eta \sim 1$ nW/Hz$^{1/2}$. This large difference is not due to clamping perimeter alone, because the drumheads only have a circumference 31 times that of the most sensitive device. Therefore, we observe an immediate improvement in sensitivity in going from the drumhead to the trampoline. Our lowest $\eta$ compares favorably to the state-of-the-art in room-temperature bolometry, competing with vanadium oxide resistive bolometers[37,38] $\sim$ 1-10 pW/Hz$^{1/2}$, other resonant microbolometers[32,33] $\sim$20-30 pW/Hz$^{1/2}$, and greatly outperforming graphene hot-electron bolometers[14] $\sim$ 500 pW/Hz$^{1/2}$.

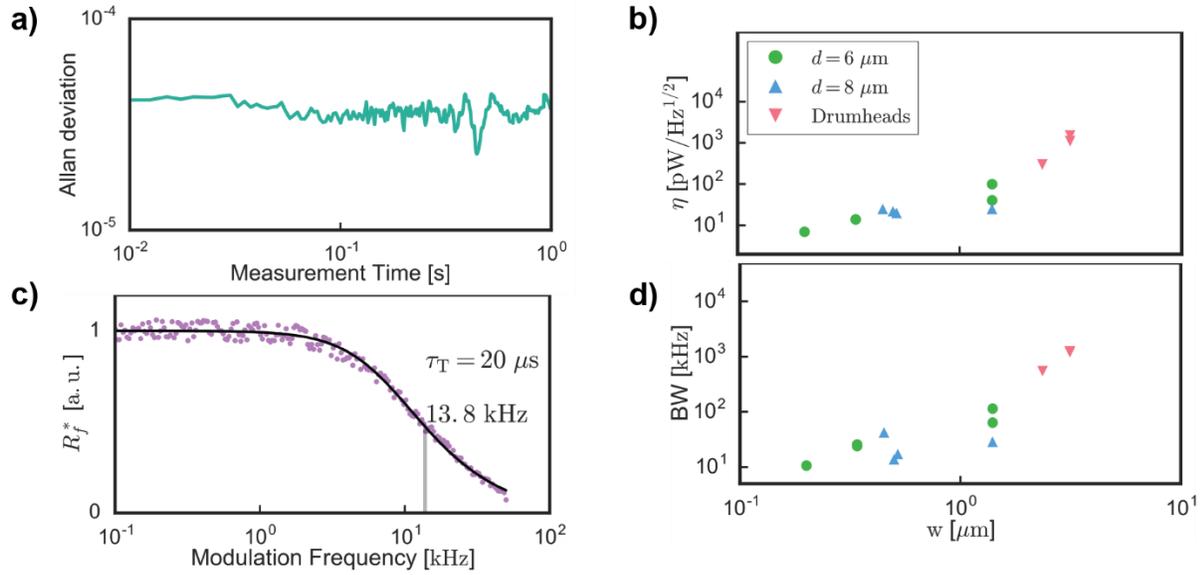

Figure 3: **a)** Allan deviation of the frequency noise vs. measurement time in a log-log plot. The resonance frequency was tracked with the PLL to obtain temporal frequency data. The Allan deviation exhibits a flat response. **b)** Sensitivity vs. tether width for 9 different trampolines and 3 different drumheads. For the drumheads, the tether width is taken to be as 1/4 of their diameter. **c)** Measured resonance shift as 40 Hz of 190 nW of peak heating radiation is applied while sweeping the modulation frequency from 100 Hz to 50 kHz. The total resonance shift was found to be constant for low modulation frequency and reached half its maximum value at 13.8 kHz. A thermal circuit model was used to fit the thermal response time of the trampoline, as described in the text. **d)** 3 dB bandwidth vs. tether width for 9 different trampolines and 3 different drumheads. For the drumheads, the tether width is taken to be 1/4 of the drumhead diameter.

The response speed of a bolometer determines its ability to detect transient signals and fast variations of the radiation intensity. We infer the speed or bandwidth from the 3 dB roll-off of the $R_f$ frequency response. To measure this response, we sweep the modulation frequency ($\omega$) of the heating laser at fixed power and measure $\Delta f_0$ versus $\omega$ by inputting the PLL signal into a second lock-in detector. For devices with bandwidths faster than the PLL, we used an off-resonant procedure to determine the bandwidth and response time[26,39] (see SI). An $R_f$ frequency spectrum for a trampoline is shown in Figure 3c. This spectrum has a nearly flat response up to its 3 dB bandwidth of 13.8 kHz, corresponding to a ~ 20 µs response time. Like the sensitivity, the response bandwidth ($BW$) increases with tether width, as shown in Figure 3d. The bandwidth of trampolines varied between 10 kHz to 100 kHz, while the drumheads have bandwidth as high as 1.3 MHz (220 ns response time), a value ~ 120 times greater than the slowest trampoline. Furthermore, the response bandwidth demonstrated by our GNB is over a 1000 times faster than modern micromechanical bolometers[33], and nearly 100,000 times faster than vanadium oxide resistive bolometers[37,38].

A common figure of merit[40] used to compare bolometers is given by the product of the incident sensitivity and response time, $FOM \propto \eta \times \tau_T$. The lowest reported $FOM$ for room-temperature microbolometers[33,37,38,40] is of order $10^5$ mK ms µm². Although it is not yet optimized and has low optical absorption (2.3%), our GNB has already matched this record-low $FOM$, nearly constant across all devices. With improvements such as increasing the absorption, using low-stress graphene, and operating at higher pressures, we expect to attain $FOM \sim 10$ mK ms μm$^2$ (see SI).

**Thermal circuit and membrane mechanics modeling**

The response bandwidth and the overall response spectrum (Figure 3c) of our GNB are well described by a thermal RC circuit model, shown schematically in Figure 4a, which predicts that smaller area devices with wider tethers or larger numbers of tethers will yield a higher $BW$. In this model, $R_T$ is the combined thermal resistance of all four tether supports, $C$ is the thermal capacitance given by the heat capacity of the suspended graphene, and $I$ is the amplitude of absorbed power modulated at frequency $\omega$. Also, the surrounding support is assumed to be a thermal sink because of its relatively large thermal mass. The model predicts a temperature difference of

$$\Delta T = I \frac{R_T}{\sqrt{1 + \omega^2 R_T^2 C^2}} \tag{5}$$

between the graphene and the surrounding support, and a characteristic time of $\tau_T = R_T C$—the thermal RC time constant.

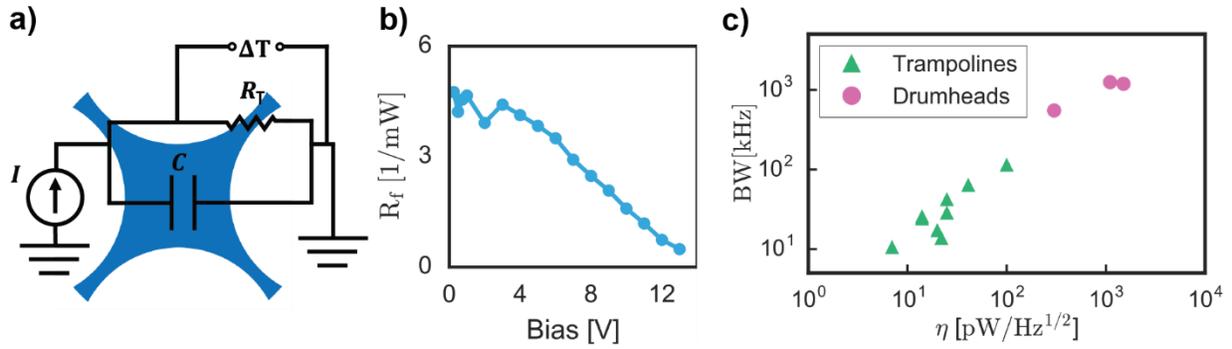

Figure 4: **a)** Schematic of the thermal circuit model used to model the bolometer performance. **b)** The frequency responsivity vs. bias voltage for the graphene trampoline. The bias voltage increases the stress on the trampoline which decreases its frequency responsivity. **c)** Frequency responsivity and thermal response time for both trampolines and high perimeter drumheads. We observe a tradeoff between sensitivity and bandwidth.

We relate $\Delta T$ to $R_f$ by examining the change in resonance frequency ($\Delta f_0$) that results from heating. According to membrane theory, $\Delta f_0 \propto \Delta \sigma$, where $\Delta \sigma$ is a change in mechanical stress, and the stress caused by thermal expansion is $\Delta \sigma \propto \Delta T$, therefore, $R_f \propto \Delta T$. We fit the model prediction for $\Delta T$ to the $R_f$ spectra and find excellent agreement, as seen in the black fit curve in Figure 3c. In this model, $\tau_T = R_T C$ defines the response bandwidth by $BW = \sqrt{3}/(2\pi R_T C)$. Taking the thermal resistance as $R_T \sim \rho_T l/w$, where $\rho_T$ is the 2D thermal resistivity, and $l$ and $w$ are the tether length and width, respectively, we have $BW \propto w$, in accord with the $BW$ data shown in Figure 3d. By using the measured $BW$ and an estimate for the heat capacity (see SI), we can calculate the thermal resistance of our devices; for the data shown in Figure 3c, $\tau_T = 20$ μs and $R_T \sim 5 \times 10^8$ K/W. This $R_T$ is exceptional and leads to high temperatures; at 1 μW, the graphene resonator will reach a temperature of nearly 1000 K.

The membrane mechanics and thermal circuit models provide insight into how material properties and device geometry influence the bolometer sensitivity. A larger $R_T$ is expected to yield more sensitive bolometers (*i.e.* a lower noise-equivalent power). From the thermal Ohm's Law ($\Delta T = IR_T$) and $R_f$, we obtain the relation $R_f = \alpha_T R_T$, where $\alpha_T = (\Delta f_0/f_0)(1/\Delta T)$ is the temperature coefficient of frequency. To first order, thermal stress in a membrane leads to

$$R_f = \frac{Y\,\alpha}{2\,\sigma_0(1-\nu)} R_T \tag{6}$$

where $Y$ is the Young's modulus, $\alpha$ is the thermal expansion coefficient, $\sigma_0$ is the initial stress, and $\nu$ is the Poisson ratio, and $\alpha_T = (Y\,\alpha)/(2\sigma_0(1-\nu))$. For a given device, Equation 6 predicts that $R_f$ will decrease with added stress in the graphene. To check this prediction, we apply electrostatic stress with a back-gate bias $V_{DC}$, which pulls the graphene structure toward the silicon back-gate, while simultaneously measuring $R_f$. As predicted, we see that $R_f$ decreases monotonically with increasing bias, as shown in Figure 4b. For fixed stress, Equation 6 also predicts $R_f \propto R_T$, and assuming a simple width dependence of the resistance, we have $R_f \propto 1/w$. The experimental data for $R_f$ (Figure 2c) agrees well with this prediction. Accordingly, the model also predicts $\eta \propto w$. For the number of devices tested in this work, the tether width dependence of $\eta$ agrees well with the linear prediction, as shown in Figure 3b, despite variations in the Allan deviation. Thus, lower stress devices with a narrower tether width will be more sensitive to light.

By examining all our bolometers, we observe that the bandwidth is proportional to the sensitivity—a faster device is also less sensitive. The data illustrating this relationship is in Figure 4c, showing trampolines are slower but more sensitive than the drumheads. Our model predicts this proportionality through the expression

$$BW = \left(\frac{\sqrt{3}}{2\pi}\frac{\alpha_T}{\sigma_A\sqrt{t}}\frac{1}{C}\right)\cdot \eta \tag{7}$$

Accordingly, for a given sensitivity, less stress and a smaller heat capacity (*i.e.* smaller device area) will boost the speed.

**Discussion**

Our bolometer technology operates in a unique parameter space characterized by excellent sensitivity, high speed, high operation temperature, large dynamic range, and small cell (pixel) size. By examining the fundamental temperature fluctuation detection limit[1,41], $\eta_{TF} = \sqrt{\frac{4\,k_B\,T^2 t}{R_T}}$, we see that the primary way to improve sensitivity is to increase the thermal resistance. Through a feasible narrowing of the tether width, we expect to attain $R_T \sim 10^{10}$ K/W while preserving $BW \sim$ kHz. In many other bolometer architectures, a thermal resistance of this magnitude would be prohibitive for several reasons. First, according to the relation $\tau_T = R_T C$ and because of the high intrinsic heat capacity of these systems, the bandwidth falls far below the 30 Hz bandwidth commonly used for imaging. In contrast, graphene has the lowest-possible heat capacity per unit area of any material, so it can possess both a high thermal resistance and a fast bandwidth.

Second, a high thermal resistance can heat the bolometer to a temperature well above its melting point (> 1000 K), leading to irreversible damage and device failure known as the "sunburn effect"[33]. To avoid sunburn, the cell size is typically made larger than 10 μm, decreasing the pixel resolution and limiting the dynamic range. However, suspended graphene has excellent thermal stability, possessing a stable nanomechanical response up to at least 1200 K[42] and a melting point well above 2600 K[43]. Therefore, our graphene bolometer is relatively immune to sunburn, and can thus benefit from both a smaller pixel size and a large dynamic range.

It is possible to greatly improve the sensitivity and speed of our graphene nanomechanical bolometers through practical modifications to material properties and device geometry. First, our devices used CVD graphene which had higher initial stress and a softened elastic modulus[44], so using exfoliated[24] or low stress[45] graphene could improve $\eta$ by a factor of 10 without sacrificing bandwidth. Although the focus of this work is room-temperature bolometry, operating at a lower temperature would lower the stress and thermal noise[1], and increase the magnitude of the thermal expansion coefficient[25], which by Eq. 5 will improve the sensitivity. Absorption could be increased to near unity by placing the bolometer in a $\lambda/4$ cavity[33], using multilayer graphene[8], or by utilizing graphene plasmonics in the mid-IR[46]. We can reduce the $1/f$ frequency noise by another order of magnitude by operating at intermediate pressures[47] and by using a more massive device[36]. The largest and simplest improvement would be to increase $R_T$. By using longer, narrower tethers (as narrow as 10 nm[28]) and by creating defects[48] in the tethers with FIB, we could increase $R_T$ by over a 100-fold. Taken together, these improvements could bring the sensitivity down to 5 fW/Hz$^{1/2}$, a value that is competitive with modern diode-based visible light photoreceivers and superconducting transition edge detectors.

The GNBs developed here are scalable and offer straight-forward integration into practical devices. These GNBs could be operated with fully-integrated electrical detection and actuation[49,50], eliminating the need for an external interferometer. Alternatively, the GNB can be used in an all-optical platform[28], eliminating the need for on-chip electronics and enabling operation in high-temperature environments. The process used to make these devices is simple, involving one single-step transfer of CVD graphene, and is compatible with chip-scale fabrication and the production of dense bolometer arrays.

**Conclusion**

We used a suspended graphene nanomechanical resonator as a new type of room-temperature light-detecting bolometer. Using our approach, we achieve a sensitivity of 7 pW/Hz$^{1/2}$ and a bandwidth over 1 MHz, thus demonstrating the feasibility of fast, sensitive, room-temperature graphene bolometry. With a few practical improvements, we expect our approach will reach the femtowatt regime without sacrificing high-speed performance. Together with its exceptional and disruptive combination of speed and sensitivity, our GNB is poised for applications that also require ultra-broadband spectral absorbance at room temperature, including terahertz biomedical imaging[3], environmental air monitoring and water detection, and high-speed scientific optical imaging[2] and spectrometry. The extreme high-temperature compatibility of our bolometer will make it useful for relevant safety and security applications, such as firefighting and industrial process monitoring, and in scientific experiments that take place at high temperature, such as close-proximity solar imaging[4]. Moreover, the thermal stability of our bolometer will enable the downsizing of the pixel size to a micrometer or smaller, which would drastically increase (~ 100X) the resolution of thermal imaging and night-vision systems.

## Acknowledgments

We acknowledge the facilities and staff from the Center for Advanced Materials in Oregon (CAMCOR), and the use of the University of Oregon's Rapid Materials Prototyping facility, funded by the Murdock Charitable Trust. We thank Joshua Ziegler, Rudy Resch, and Kara Zappitelli for scientific discussions and feedback related to this work. This work was supported by the University of Oregon and the National Science Foundation (NSF) under grant No. DMR-1532225.

## Author Contributions

AB and BA conceived and designed the experiments. DM and AB fabricated the graphene devices. DM designed and built optical measurement apparatus with assistance from AB. AB performed the experiments and analyzed the data. AB and BA wrote the paper, with feedback from DM. BA supervised the work.

# Methods

# A fast, sensitive, room-temperature graphene nanomechanical bolometer

Andrew Blaikie, David Miller, Benjamín Alemán

Department of Physics, University of Oregon, Eugene, Oregon 97403
Material Science Institute, University of Oregon, Eugene, Oregon 97403
Center for Optical, Molecular, and Quantum Science, University of Oregon, Eugene, Oregon 97403

**Fabrication of Silicon Devices**

We fabricated suspended graphene mechanical resonators using standard semiconductor processing techniques. We began by growing 1 µm of wet thermal oxide on $Si^{++}$ wafers at 1100 C. Next, we patterned 6-8 µm diameter holes with AZ1512 photoresist and a direct write laser photolithography system. We etched 600 nm deep into the oxide with a dry inductively coupled plasma etch using a plasma of $CHF_3$ and argon. By leaving some of oxide intact, any collapsed graphene could not cause a short between the suspended graphene and the $Si^{++}$. We then patterned metal electrodes using another AZ1512 direct write photolithography step. Next, we evaporated 5/50 nm Ti/Pt using electron beam evaporation followed acetone liftoff with sonication.

**Semi-Dry Graphene Polymer Transfer**

A semi-dry polymer supported transfer technique was used to place a large sheet of commercial monolayer graphene on Cu foil (Graphenea) over the exposed holes and metal contacts according to the procedures outlined by Suk et. al[1]. First, a ~3 micron thick layer of PMMA A11 was spun onto the Graphene/Cu. The graphene on the backside of the foil was removed with oxygen plasma. Then, a 1 mm thick piece of PDMS with a ~1 cm diameter hole punched through the middle of it was placed on top of the Graphene/Cu stack. A thin plastic backing was left on the PDMS to increase the rigidity of the film. The Cu foil was etched on a solution of ammonium persulphate (40 mg/ml). The relatively rigid PDMS/PMMA/Gr stack was picked up with tweezers and placed in three sequential water baths before being removed and dried in air. Concurrently, the target substrate with holes was prepared by cleaning it in oxygen plasma before placing it on a hot plate at 155 C. The now dry PDMS/PMMA/Graphene stack was placed on top of the hot substrate with the through hole covering the entirety of the chip. The substrate was left for ~16 hours to improve adhesion between the graphene and the $SiO_2$. The PDMS was then peeled away and the PMMA was removed in flowing Ar and $H_2$ at 350 C for 3 hours. Graphene was scratched off the perimeter of the substrate to prevent shorting to the $Si^{++}$ gate.

**Focused Ion Beam Cutting of Trampolines**

We shaped the graphene into trampolines with a focused ion beam[2]. FIB shaping was performed in an FEI Helios 600i SEM-FIB with a Ga$^+$ source. The ion beam current and voltage were 1.1 pA and 30 kV, respectively. To fabricate a trampoline, four circle outlines were cut into a graphene drumhead using a single beam pass and a dwell time of 1 ms, which was sufficient to etch a line completely through the suspended graphene sheet. The high tension in the graphene sheet causes the graphene inside the circular cut to pull away from the trampoline resonator and collapse into the cavity. The FIB fabrication technique has a yield of near 100%, with device failures typically due to holes or other defects present in the graphene prior to milling. Although the FIB milling likely induces additional disorder in the graphene sheet, it still maintains its excellent electrical, mechanical, and thermal properties.

**Optical Measurements**

Mechanical motion was measured with optical interferometry with a 633 nm HeNe laser. All measurements were performed at room temperature under a vacuum of $P < 10^{-5}$ Torr. Measurement laser powers were kept less than one microwatt to minimize any heating caused by the measurement. Reflected light was measured with a silicon avalanche photodiode and recorded with a lock-in amplifier. The lock-in amplifier was used to apply an AC voltage to actuate motion in the suspended graphene resonators. A built in phase-locked loop and could track changes to resonant frequencies due to radiation induced heating or inherent frequency fluctuations. Heating radiation was applied with a 532 nm diode laser modulated with an acousto-optic modulator. See SI for more details.

# Supplementary Information

# A fast, sensitive, room-temperature graphene nanomechanical bolometer

Andrew D. Blaikie, David Miller, Benjamín J. Alemán

Department of Physics, University of Oregon, Eugene, Oregon 97403
Material Science Institute, University of Oregon, Eugene, Oregon 97403
Center for Optical, Molecular, and Quantum Science, University of Oregon, Eugene, Oregon 97403

**Devices Characterization**

We characterized 12 devices for this work. Figure 1 shows a gallery of SEM images of all the graphene nanomechanical bolometers (GNB) characterized. Table 1 shows the mechanical and bolometric properties of these devices.

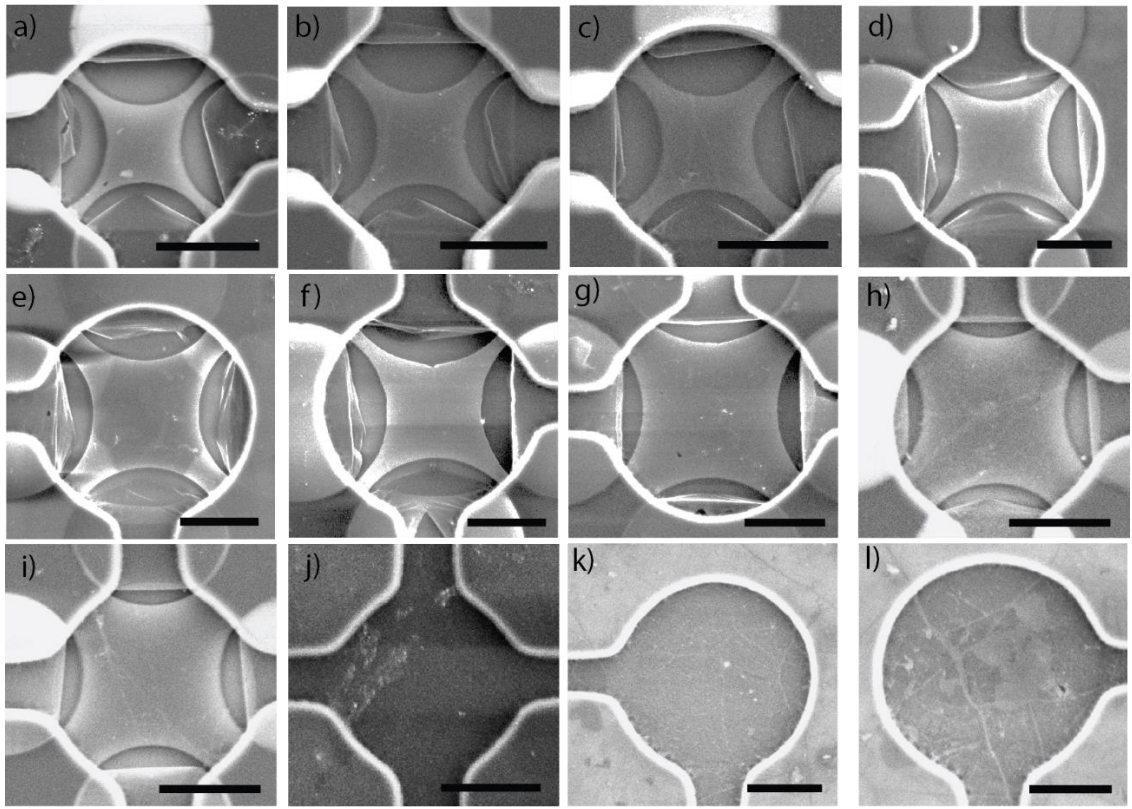

Figure 1: SEM images gallery of all devices characterized. Devices are labeled by letters a-l. Scale bar is 3 µm.

|   | $\eta$ (pW/Hz$^{1/2}$) | $R_f$ (1/W) | $\tau_T$ (µs) | $\sigma_A$ (10$^{-5}$) | $f_0$ (MHz) | $w$ (µm) | $d$ (µm) |
|---|---|---|---|---|---|---|---|
| a | 7 | 300,000 | 26 | 2.1 | 16.6 | 0.20 | 6 |
| b | 14 | 100,000 | 10.7 | 1.4 | 21.8 | 0.34 | 6 |
| c | 14 | 93,000 | 11.4 | 1.3 | 25.0 | 0.34 | 6 |
| d | 20 | 210,000 | 16 | 4.3 | 9.6 | 0.52 | 8 |
| e | 22 | 180,000 | 20 | 4.1 | 10.7 | 0.50 | 8 |
| f | 25 | 260,000 | 22 | 6.5 | 11.1 | 0.45 | 8 |
| g | 25 | 98,000 | 9.6 | 2.5 | 11.0 | 1.4 | 8 |
| h | 41 | 39,000 | 4.3 | 1.6 | 25.4 | 1.4 | 6 |
| i | 100 | 25,000 | 2.4 | 2.7 | 24.0 | 1.4 | 6 |
| j | 300 | 7,500 | 0.5 | 2.3 | 21.9 | - | 6 |
| k | 1100 | 3,700 | 0.22 | 4.2 | 8.1 | - | 8 |
| l | 1500 | 2,600 | 0.23 | 3.8 | 11.4 | - | 8 |

Table 1: Shows the detector sensitivity ($\eta$) at a 100 Hz bandwidth, frequency responsivity ($R_f$) to absorbed power, thermal response time ($\tau_h$), Allan Deviation ($\sigma_A$) over a 10 ms integration time, initial resonance frequency ($f_0$), tether width ($w$), and initial diameter ($d$), for the 12 bolometers studied in this work.

## Interferometric Transduction of Mechanical Motion

The motion of the graphene mechanical resonators was transduced with optical interferometry and lock-in amplification. Motion was actuated with a combination of a $V_{DC}$ and $V_{AC}$ electrical bias[1] between the graphene and the Si$^{++}$ which produces a drive force, $F_D \propto V_{DC} V_{AC} \cos \omega t$. A 633nm probe laser (<1 µW) was focused down onto the graphene trampoline using a 40x, 0.6 NA objective. A low-finesse Fabry-Perot cavity, formed between the Si$^{++}$ and the graphene, applies a small modulation to the reflected light as the resonator vibrates. We used a polarizing beam splitter and a quarter waveplate to split the reflected light from the incident beam. The intensity of the reflected beam was converted to a voltage using a silicon avalanche photodiode before being fed into a lock-in amplifier referenced to the applied $V_{AC}$ electrical drive signal. The mode shape of the graphene drums[2] could be visualized by scanning the 633 nm probe laser across the device using a fast steering mirror with diffraction limited resolution. All measurements were done under vacuum at less than 10$^{-5}$ Torr to minimize air damping. This optical setup is shown in Figure 2.

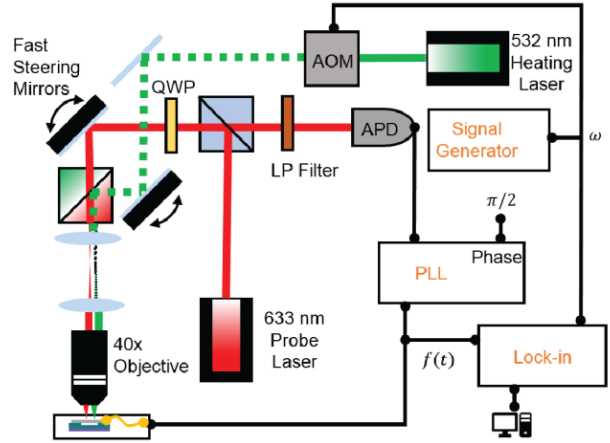

Figure 2: Sketch of the optical interferometer setup used to measure the motion of and apply heating radiation to the suspended graphene.

## Bolometric Measurements

To apply heating radiation, we used a 532 nm laser modulated with an acousto-optic modulator (AOM) with a sine wave. A dichroic beamsplitter was used to couple the heating laser into the optical path. We measured the incident power with a Thorlabs S120VC optical power meter just before entering the objective and estimate that the absorbed power is 2.3% of this value. When the suspended graphene absorbs light its mechanical resonance shifts. To track this resonance shift, we used frequency modulation with a phase-locked-loop (PLL). The suspended graphene was electrically driven on resonance and the phase between the drive signal and amplitude signal was detected with a Zurich HFLI2 lock-in amplifier. The PLL feeds back on any deviation in phase by adjusting the drive frequency to keep this phase constant. The feedback bandwidth setting of the PLL was typically set between 1-50 kHz to not limit the frequency response. By reading out the adjusted drive frequency time series data, the PLL tracks any changes to mechanical resonance induced by the absorption of light or inherent frequency fluctuations. This frequency time series data was then fit to a sine wave with a frequency matching the AOM modulated frequency of the heating laser. The amplitude of this fitted sine wave was used to calculate the frequency shift responsivity. Measurements of frequency noise were also performed with the PLL in the absence of heating radiation. This frequency noise time series data was used to calculate the Allan deviation.

**Bandwidth Measurements**

We measure the bandwidth of the GNB by increasing the modulation frequency of the heating laser and monitoring the response. While tracking the resonant frequency shifts with the PLL, we output a voltage proportional to the frequency shift from the PLL and input this into a second lock-in amplifier channel in the Zurich HFLI2. This signal is referenced to input of the AOM. By sweeping the frequency of the AOM drive signal we could quickly extract how the resonant frequency shift amplitude drops as the modulation frequency of the heating laser increases. This data was used to fit the thermal response time and was used to estimate the 3dB bandwidth.

For GNB that could respond faster than ~30 kHz we used an off-resonant method to estimate the thermal response time. For these cases, we were unable to measure the intrinsic bandwidth and thermal response time by looking at the change in frequency shift with the phase locked loop (PLL), because the bandwidth of the PLL is unable to track changes in resonance frequency at high frequencies. For these cases, we infer $\tau_T$ using thermomechanics[3,4]. For this measurement, we modulate the heating laser at frequencies below mechanical resonance without applying any electrical actuation. In this regime, the mechanical amplitude is assumed to be proportional to the change in temperature, $A \propto \Delta T$, by first order thermal expansion. Solving the thermal circuit model predicts that the amplitude will have the form,

$$A \propto \frac{1 - i\omega\tau_T}{1 + \omega^2\tau_T^2}. \tag{1}$$

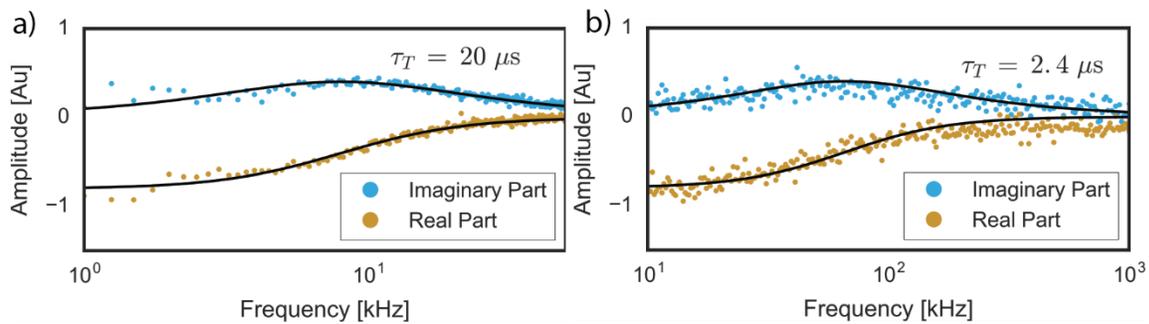

Figure 3: Real and imaginary amplitude of thermal expansion induced displacement for two trampolines with different tether widths. These were fit to the thermal circuit model to extract the thermal response time, $\tau_T$. **a)** Corresponds to device e in Figure 1. **b)** Corresponds to device i in Figure 1.

The 532 nm heating laser was used to create an AC heat source and the deflection of the graphene was measured with the 633 nm interferometer. The real and imaginary amplitudes (defined by the phase difference between the mechanical amplitude and the heating laser intensity) are measured with respect to this heating laser as shown in Figure 3 for two different GNB. The model was fit to the data to extract the thermal response time $\tau_T$. We also confirmed that this method predicts the nearly the same thermal response time as the frequency shift method, with both methods predicting $\tau_T = 20$ μs for device e in Figure 1.

**Estimation of the Mass and Heat Capacity**

We can estimate the thermal resistance, $R_T = \Delta T/I$, of the GNB using the measured thermal response time $\tau_T = R_T C$, if we know the heat capacity. The change in mechanical resonance frequency with DC bias voltage can be used to estimate the mass density of graphene drumheads[5], which we use to then estimate the heat capacity. By fitting the resonance with respect to bias voltage to the equation

$$(f^2 - c_3 V_g^2 - c_1)(f^2 - c_3 V_g^2)^2 = c_2 V_g^4, \tag{2}$$

where $c_1, c_2, c_3$ are fitting parameters, we can estimate the mass density. In terms of theory based on the electrostatic force of a parallel plate capacitor[5],

$$c_1 = \frac{2.404^2}{4\pi^2 a^2} \frac{\sigma_0}{\rho}, \tag{3}$$

$$c_2 = \frac{2.404^6 \epsilon_0^2}{6144\ \pi^6 a^4 d^4 \rho^3} \frac{Y}{1-\nu}, \tag{4}$$

and $c_3$ is a parameter used to account for capacitive softening[6]. Here $\sigma_0$ is the initial tension in the drumhead, $\rho$ is the 2D mass density, $a$ is the drumhead radius, $\epsilon_0$ is the permittivity of free space, $d = 1$ μm is the distance between the graphene and silicon, $Y = 60$ N/m the Young's modulus of CVD graphene[7], $\nu = 0.16$ is the Poisson ratio for graphene. We performed this measurement on several graphene drumheads to estimate the amount of contaminating mass on the devices. A typical measurement is shown in Figure 4 for a 6 μm graphene drumhead.

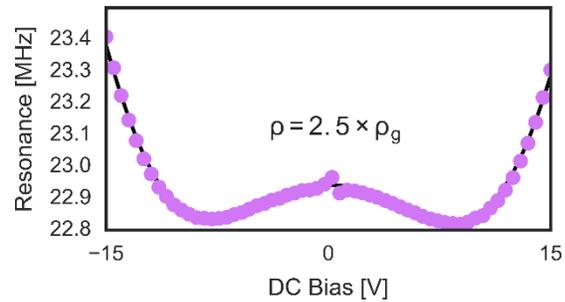

Figure 4: The resonance frequency plotted vs. bias voltage for a typical 6 μm drumhead. To extract the resonance frequencies, the data was fit using a damped driven oscillator model at varying gate voltages. Using these resonance values, the mass density for the drumhead was extracted from a fit to the theory based on the electrostatic force of a parallel plate capacitor.

To estimate the heat capacity and mass density of the trampolines we assume that the mass density for the trampolines is the same as the mass density of the drumheads. For these devices, we use a typical fitted value,

$\rho = 2.5 \times \rho_g$, where $\rho_g$ is the intrinsic mass density of monolayer graphene. This amount of contaminating mass is consistent with other graphene nanomechanical systems that used a PMMA transfer technique to suspended graphene sheets[5,8].

Next, we determine the heat capacity of the graphene nanomechanical bolometers. We assume that the added mass is due to PMMA contamination from the dry polymer transfer technique[8]. The heat capacity for the GNB is then $C = (c_g + 1.5 c_p)\rho \times a^2$, where $a^2 \sim 15$ μm$^2$ is the device area, $c_g = 700$ J/(kg K) is the specific heat of graphene, and $c_p \sim 1500$ J/(kg K) is the specific heat of PMMA. Using these values we find a heat capacity of $C \sim 3\times 10^{-14}$ J/K. Using this value the highest thermal resistance is estimated to be $R_T \sim 8 \times 10^8$ K/W for the graphene bolometers.

**Modeling the Temperature Coefficient of Frequency**

We derive a formula for the temperature coefficient of frequency, $\alpha_T = \frac{\Delta f_0}{\Delta T\, f_0}$, which illuminates how improve the noise-equivalent power because $\eta \propto \alpha_T$. The equation for the mechanical resonance for a thin circular membrane is given by

$$f_0 = \frac{4.808}{4\pi r}\sqrt{\frac{\sigma}{\rho}}, \qquad (5)$$

where $r$ is the radius, σ is the in plane stress, and ρ is the 2D mass density. When the temperature of the suspended membrane increases, the stress changes according to the relation by the stress-strain relation, $\Delta\sigma = -(\alpha\,\Delta T)\,Y/(1-v)$, where $\alpha$ is the thermal expansion coefficient[9]. Using these equations and a first order expansion we calculate the temperature coefficient of frequency ($\alpha_T$), the relative change in resonance per Kelvin to be

$$\Delta f = \frac{4.808}{4\pi r}\sqrt{\frac{\sigma - \Delta\sigma}{\gamma\rho}} - f_0 = f_0\left(\sqrt{1 - \frac{\Delta\sigma}{\sigma_0}} - 1\right) \approx -f_0\left(\frac{\Delta\sigma}{2\sigma}\right), \qquad (6)$$

$$\alpha_T = \frac{\Delta f/f_0}{\Delta T} = -\frac{\alpha Y}{2\,\sigma_0(1-v)}. \qquad (7)$$

Therefore, the noise-equivalent power can be reduced by using low-stress, high modulus, and high thermal expansion sheets of graphene.

**Figure of Merit Calculation**

The figure of merit is commonly used to compare different bolometers[10], calculated using $FOM = NETD \times \tau_T \times A_p$ where $NETD$ is the noise equivalent temperature difference and $A_p$ is the detector area[11]. The noise equivalent temperature difference is proportional to the noise-equivalent power through the relation[11]

$$NETD = \frac{4F^2}{\pi\, A_p \left(\frac{\Delta L}{\Delta T}\right)}\eta_p/\sqrt{t},$$

where $F$ is the optical aperture (typically $F = 1$), $\Delta L / \Delta T = 0.84$ W/m²/sr/K is the luminance variation with scene temperature around 300K, $\eta_p$ is the noise equivalent power to incident radiation, and $t$ is the measurement time. Increasing the thermal resistance improves $\eta_p$ at the expensive of bandwidth and increasing the detector area improves the $NETD$ at the expensive of pixel pitch. Therefore, this figure of merit removes geometric considerations when comparing bolometer technologies because both the thermal resistance and the pixel area can usually be tuned by changing the geometry. This is true for our GNB, as it is common to fabricate suspended graphene sheets with diameters ranging from 1-25 µm[12] and we have demonstrated that the thermal resistance can be tuned by varying the trampoline tether width. Doing this calculation for the most sensitive trampoline, with $\eta_p = 300$ pW/Hz$^{1/2}$, $t = 10$ ms, $\tau_T = 26$ µs, predicts $1.18 \times 10^5$ mK ms µm². The best $FOM$ obtained was with a 6 µm drumhead, which yielded $0.98 \times 10^5$ mK ms µm².

resonators. *Nano Lett.* **11,** 1232–1236 (2011).